\documentclass[times,doublespace,12pt]{article}

\usepackage{amssymb}
\usepackage{amsmath}
\usepackage{amsthm}
\usepackage{natbib}
\usepackage{graphicx}
\usepackage{verbatim}
\usepackage{placeins}

\usepackage{algorithm}
\usepackage{algorithmicx}

\usepackage{subcaption}
\usepackage{float}
\usepackage{natbib}
\bibpunct{(}{)}{;}{a}{,}{,}

  \title{Subgroup Identification using Covariate Adjusted Interaction Trees}
  \author{Jon Arni Steingrimsson and  Jiabei Yang \\
  Department of Biostatistics, Brown University}
  \date{}

\begin{document}
  \maketitle

\noindent \textbf{Abstract:} We consider the problem of identifying sub-groups of participants in a clinical trial that have enhanced treatment effect. Recursive partitioning methods that recursively partition the covariate space based on some measure of between groups treatment effect difference are popular for such sub-group identification. The most commonly used recursive partitioning method, the classification and regression tree algorithm, first creates a large tree by recursively partitioning the covariate space using some splitting criteria and then selects the final tree from all subtrees of the large tree. In the context of subgroup identification, calculation of the splitting criteria and the evaluation measure used for final tree selection rely on comparing differences in means between the treatment and control arm. When covariates are prognostic for the outcome, covariate adjusted estimators have the ability to improve efficiency compared to using differences in means between the treatment and control group. This manuscript develops two covariate adjusted estimators that can be used to both make splitting decisions and for final tree selection. The performance of the resulting covariate adjusted recursive partitioning algorithm is evaluated using simulations and by analyzing a clinical trial that evaluates if motivational interviews improve treatment engagement for substance abusers.

\section{Introduction}

Identifying sub-groups of participants enrolled in a clinical trial that have enhanced treatment effects helps focus future trials on participants that are more likely to benefit from treatment, leading to potentially more targeted treatment therapies. It is also important that the method used for sub-group identification correctly identifies settings with no such sub-groups. Sub-group identification is traditionally done by testing for treatment and covariate interactions in a generalized linear model (GLM), but the use of more data driven methods for subgroup identification is becoming more popular. \citet{lipkovich2017tutorial} provide a recent review of data driven subgroup identification.

Due to its ability to detect interactions, the classification and regression tree (CART) algorithm  \citep{breiman1984classification} has been modified by several authors to detect sub-groups with enhanced treatment effects. The CART algorithm recursively partitions the covariate space using some splitting criteria. \citet{foster2011subgroup} use the random forest algorithm to predict individual treatment effects for all observations and run a CART tree on the predictions. For repeated outcomes, \citet{su2011interaction} proposed to base splitting decisions in the tree building process on a Wald tests statistic for a treatment and covariate interaction from a generalized estimating equation model. \citet{seibold2016model} used a model based recursive partitioning method for sub-group identification where decisions are based on the instability of parameter estimation. Finally, the Interaction Trees algorithm \citep{su2009subgroup} modifies the CART algorithm by making splitting decisions based on differences between average treatment effect for each node using node specific means for each treatment arm. 

When the covariates are prognostic for the outcome, covariate adjusted estimators can substantially improve efficiency over calculating the mean of the outcome restricted to each treatment arm \citep{zhang2008improving,steingrimsson2017improving}. A simple covariate adjusted estimator is the model standardization estimator first proposed by \citet{scharfstein1999adjusting}. Implementation of the model standardization estimator requires modeling the relationship between the outcome, a treatment indicator, and the covariates. When modeled using a GLM, the model standardization estimator is consistent even if the GLM is misspecified. If the model is correctly specified, it is semi-parametric efficient and more efficient than using simple means. 

This manuscript proposes to replace node specific means which govern the decision making for the Interaction Trees algorithm by the more efficient covariate adjusted estimator. We refer to this new algorithm as the covariate adjusted interaction tree (CAIT) algorithm. We furthermore derive a new class of covariate adjusted estimators for the node specific treatment effect. This class improves upon the model standardization estimators in two ways. First, it allows the use of more flexible outcome models than a GLM while remaining consistent under model mis-specification. Second, the outcome model can be fit using the whole dataset rather than using only the data falling in the node. This leads to potential variance reduction as well as computational savings as the outcome model only needs to be fit once. 

Section \ref{sec:CA-Trees} defines the CAIT algorithm and introduces a novel method for final tree selection. Section \ref{sec:CA-Est} develops and derives properties of the covariate adjusted node specific estimators. Section \ref{sec:Simulations} evaluates the performance of the CAIT algorithm using simulations. Finally, Section \ref{sec:Analysis} presents an analyzes of a clinical trial evaluating if motivational interviews improve treatment engagement for substance abusers.

\section{Covariate Adjusted Interaction Trees}
\label{sec:CA-Trees}

Let $Y$ be an outcome which can be either binary, continuous, or counts. Let $A$ be an indicator if participant is randomized to the treatment arm ($A=1$) or to the control arm ($A=0$). Let $X$ be a vector of baseline covariates measured prior to randomization taking values in $\mathcal{X}$. Following \citet{seibold2016model}, we define a covariate to be prognostic if it is predictive of the outcome and a covariate is said to be predictive if it is predictive of the treatment effect. A set $w$ is called a subgroup if $w$ is a subset of $\mathcal{X}$. The data collected is assumed to consists of $n$ i.i.d.replications of $(Y,X,A)$ where $X$ is bounded and $E[Y] < \infty$. The treatment randomization ensures that $A$ is independent of $X$, but unless otherwise stated we put no other restrictions on the joint distribution of $(Y,X,A)$.

For $l=0,1$ and a subgroup $w$, let $\hat \mu_{l}(w)$ be an estimator for $\mu_{l}(w) = E[Y|A=l, X \in w]$. Section \ref{sec:CA-Est} describes three different estimators for $\mu_{l}(w)$. The CAIT algorithm consists of the following steps:
\\
\\
\textit{1. Creating a Maximum Sized Tree:} At the beginning of the tree building process all observations are in a single node. Let $X^{(j)}$ be the j-th component of the covariate vector $X$. For a given $c \in \mathbb{R}$, split the covariate space into two groups $L = \{X^{(j)} < c\}$ and $R = \{X^{(j)} \geq c\}$. The group specific average treatment effect estimators are $\hat \mu_{1}(L) - \hat \mu_{0}(L)$ and $\hat \mu_{1}(R) - \hat \mu_{0}(R)$ for groups $L$ and $R$, respectively.
A test statistic for the difference between the treatment effects in the left and right groups is given by
\begin{equation}
\label{split-stat-2}
\left(\frac{(\hat \mu_{1}(L) - \hat \mu_{0}(L)) - (\hat \mu_{1}(R) - \hat \mu_{0}(R))}{\sqrt{\hat \sigma_1(L) + \hat \sigma_0(L) + \hat \sigma_1(R) + \hat \sigma_0(R)}} \right)^2,
\end{equation}
where $\hat \sigma_l(L), \hat \sigma_l(R), l = 0,1$ are estimators for $Var(\hat \mu_l(L))$ and $Var(\hat \mu_l(R))$, respectively. If the treatment effect is identical in both groups, the splitting statistic \eqref{split-stat-2} converges to a $\chi^2(1)$ distribution. When $\hat \mu_{l}(w), l \in \{0,1\}, w \in \{L, R\}$ are group specific means and $\hat \sigma_l(w), l \in \{0,1\}, w \in \{L, R\}$ are the pooled variance estimators, then \eqref{split-stat-2} reduces to the splitting statistic used in \citet{su2009subgroup}.

To split the node into two new nodes the CAIT algorithm cycles through all covariate and split-point combinations $(X^{(j)},c)$ and selects the pair that results in the largest value of the splitting statistic \eqref{split-stat-2}. This process is repeated within each new node until some pre-determined criteria are met. This results in a large tree denoted $\psi_{Max}$.

The above description assumes that $X^{(j)}$ is continuous. If $X^{(j)}$ is categorical, the algorithm is modified to search through all possible combinations of levels of $X^{(j)}$ for nominal covariates and all possible splits that preserve the ordering of groups for ordinal variables.
\\ 
\\ 
\noindent \textit{2. Pruning:} The pruning step creates a finite sequence of candidate trees. The following cost complexity pruning algorithm was developed in \citet{leblanc1993survival} and adapted to the setting of subgroup identification in \citet{su2009subgroup}. For a given penalization parameter $\lambda$, define the split complexity for a tree $\psi$ as
\begin{equation}
\label{split-complexity}
G^{(\lambda)}(\psi) = \sum_{i \in I_\psi} G_i(\psi) - \lambda |I_{\psi}|.
\end{equation}
Here, $I_\psi$ is the set of internal nodes of $\psi$ and $G_i(\psi)$ is the value of the splitting statistic for internal node $i$ in tree $\psi$. 

Weakest link pruning is an algorithm which creates a finite sequence of subtrees of $\psi_{Max}$ by cutting the ``weakest link'' based on the split complexity. For a non-terminal node $k$ of a tree $\psi$, define $\psi^*_k$ as the tree rooted at node $k$. That is, $\psi^*_k$ consists of node $k$ and all descendants of node $k$. The split complexity of $\psi^*_k$ is $\sum_{i \in I_{\psi^*_k}} G_i(\psi^*_k) - \lambda |I_{\psi^*_k}|$. The split-complexity of $\psi^*_k$ is zero when $\lambda = \sum_{i \in I_{\psi^*_k}} G_i(\psi^*_k)/|I_{\psi^*_k}|$. This is the cut-off where removing branch $\psi^*_k$ becomes preferred to keeping $\psi^*_k$, measured in terms of having larger split complexity.

The weakest link pruning algorithms cycles through all non terminal nodes and creates a sequence of subtrees of $\psi_{Max}$ using the following steps:
\begin{enumerate}
\item Set $\psi_0 = \psi_{Max}$ and $k = 0$.
\item Define the function $g(h) = \sum_{i \in I_{\psi^*_h}} G_i(\psi^*_h)/|I_{\psi^*_h}|$ if $h \in I_{\psi_k}$ and $g(h)=+\infty$ otherwise. The weakest link of the tree $\psi_k$ is the node $h^* = \min_{h' \in I_{\psi_k}} g(h')$. Define $\psi_{k+1}$ as the subtree of $\psi_k$ with branch $\psi^*_{h^*}$ removed. Set $k = k+1$.
\item Repeat Step 2 until $\psi_{k+1}$ consists only of the root node.
\end{enumerate}
Weakest link pruning results in a sequence of nested trees $\psi_{M}, \ldots, \psi_1, \psi_0 = \psi_{Max}$.
\\ 
\\
\noindent \textit{3. Final Tree Selection:} This step selects the final tree from the sequence of trees generated during the pruning step. The first tree selection method is an adaptation of the pruning method described in \citet{leblanc1993survival}. The training set is split into an initial tree building dataset and a validation dataset. The sequence of candidate trees built using step 1 and 2 is calculated using only the initial tree building data. For a candidate tree $\psi_k, k \in \{0, \ldots, M\}$, the value of the splitting statistic $G_i(\psi_k), i \in I_{\psi_k}$ in equation \eqref{split-complexity} is re-calculated using the validation sample. The re-calculated $G_i(\psi_k)$ is used to calculate the validation set split complexity of tree $\psi_k$ using formula \eqref{split-complexity}. 

The final tree from the sequence $\psi_{M}, \ldots, \psi_1, \psi_{Max}$ is the tree which maximizes the validation split complexity for a fixed penalization parameter. A common criterion for selecting the penalization parameter is some quantile of the asymptotic distribution of the split statistic.

Now we describe an alternative method for final tree selection which is more closely aligned with the cross-validation approach used for the original CART algorithm. The main difficulty with directly extending the cross-validation approach of \citet{breiman1984classification} is that the treatment effect is not observed on any participant. 

To overcome this difficulty, we propose the following novel cross-validation procedure. For a given split into a training and a test set and a candidate tree $\psi_k, k \in \{1, \ldots, M\}$, re-estimate the terminal node estimators using only the training data falling in each terminal node. Use tree $\psi_k$ with the re-estimated terminal node estimators to predict the treatment effect for all participants in the test set and refer to the predictions as the CAIT test set predictions. Estimate $E[Y|A, X]$ using the random forest algorithm \citep{breiman2001random} fit to the training data. Use the random forest fit to calculate prediction for the treatment effect $E[Y|A=1, X] - E[Y|A=0, X]$ for all test set participants. Calculate the cross-validation error for tree $\psi_k$ corresponding to this particular split into test and training set as the average $L_2$ distance between the CAIT test set predictions and the random forest treatment effect test set predictions. The final tree is selected as the tree that results in the smallest cross-validation error averaged over all splits into test and training sets. This tree selection method is motivated by the random forest algorithm resulting in a more flexible prediction model compared to the CART algorithm and usually having substantially better prediction accuracy.

In the simulations and data analysis presented in Sections \ref{sec:Simulations} and \ref{sec:Analysis} we refer to the first final tree selection method as final tree selection method 1 and the second final tree selection method as final tree selection method 2. 

\section{Covariate Adjusted Estimators for Node Specific Means}
\label{sec:CA-Est}


The unadjusted estimator for treatment $l \in \{0,1\}$ in group $w$ is $\hat \mu_{(Unad,l)}(w) = \frac{1}{n_l(w)} \sum_{i=1}^n I(X_i \in w) I(A_i = l) Y_i$. Here, $n_l(w) = \sum_{i=1}^n I(X_i \in w) I(A_i = l)$ is the number of participants in subgroup $w$ that are assigned to treatment arm $l$. The estimator $\hat \mu_{(Unad,l)}(w)$ is simply the average of the outcome restricted to the participants that are assigned to treatment $l$ in group $w$. Under the assumptions stated at the beginning of Section \ref{sec:CA-Trees}, $\hat \mu_{(Unad,l)}(w)$ is a consistent estimator for $\mu_l(w)$. Apart from being restricted to participants in group $w$, $\hat \mu_{(Unad,l)}(w)$ does not use information from $X$. Now we describe two covariate adjusted estimators for $\mu_l(w)$ that leverage information in $X$.

Define the GLM with a canonical link function $g(\cdot)$ as
\begin{equation}
\label{GLM-Mod}
g(E[Y|A,X; \beta]) = \beta_0 +  \beta_1 A + \beta_2^T X.
\end{equation}
The estimator $\hat \beta = (\hat \beta_0, \hat \beta_1, \hat \beta_2^T)^T$ for the regression coefficient $\beta = (\beta_0, \beta_1, \beta_2^T)^T$ restricted to group $w$ is calculated by solving
\begin{equation}
\label{Est-Eq-1}
\sum_{i=1}^n I(X_i \in w) (Y_i - h(\beta_0 + \beta_1 A_i + \beta_2^T X_i)) (1, A_i, X_i^T)^T = 0,
\end{equation}
where $h(\cdot) = g^{-1}(\cdot)$. The population quantity $\tilde \beta$ that $\hat \beta$ consistently estimates satisfies
\begin{equation}
\label{Est-eq-2}
E[I(X \in w) (Y - h(\tilde \beta_0 + \tilde \beta_1 A + \tilde \beta_2^T X)) (1,A, X^T)^T] = 0,
\end{equation}
even if model \eqref{GLM-Mod} is mis-specified. Define the covariate adjusted estimator 
\begin{equation}
\label{EstMS}
\hat \mu_{(MS, l)}(w) = \frac{1}{n(w)} \sum_{i=1}^n I(X_i \in w) h(\hat \beta_0 + \hat \beta_1 l + \hat \beta_2^T X_i),
\end{equation}
where $n(w) = n_1(w) + n_0(w)$. The estimator \eqref{EstMS} is referred to as the model standardization estimator. Here, $h(\hat \beta_0 + \hat \beta_1 l + \hat \beta_2^T X)$ is a prediction for $E[Y|A=l, X]$ from the GLM \eqref{GLM-Mod}. Hence, $\hat \mu_{(MS, l)}(w)$ averages over the prediction for all subjects in group $w$ setting their treatment assignment to $l$. Importantly, the average is taken over all subjects in group $w$ not just those assigned to treatment $l$. 

Supplementary Web Appendix \ref{sec:consistency} shows that $\hat \mu_{(MS,l)}(w)$ is a consistent estimator for $E[Y|A =l, X \in w]$ even if the GLM is misspecified. Furthermore, $\hat \mu_{(MS,l)}(w)$ is locally efficient meaning that if the GLM is correctly specified $\hat \mu_{(MS,l)}(w)$ is asymptotically efficient. In particular, this implies that if the GLM is correctly specified $\hat \mu_{(MS,l)}(w)$ is asymptotically at least as efficient as $\hat \mu_{(Unad, l)}(w)$ \citep{rosenblum2016matching}. A variance estimator for $\hat \mu_{(MS,l)}(w)$ is given by Equation \eqref{Var-1} in Supplementary Web Appendix \ref{sec:consistency}.

In the special case of linear regression ($Y$ is continuous and $g(x) = x, \forall x \in \mathcal{X}$), the model standardization estimator for the treatment effect $\hat \mu_{(MS, 1)}(\mathcal{X}) - \hat \mu_{(MS, 0)}(\mathcal{X}) = \hat \beta_1$. \citet{yang2001efficiency} showed that $\hat \beta_1$ is asymptotically at least as efficient as $\hat \mu_{(Unad, 1)}(\mathcal{X}) - \hat \mu_{(Unad, 0)}(\mathcal{X})$ under arbitrary misspecification of the linear model. When the GLM is non-collapsible, e.g.~a logistic regression model, the model standardization estimator is not equal to the estimated coefficient associated with the treatment arm from the GLM, with the former estimating a marginal effect and the latter a conditional effect. 

The consistency of $\hat \mu_{(MS,l)}(w)$ under mis-specification of \eqref{GLM-Mod} relies on using a GLM to estimate $E[Y|A=l,X]$. The efficiency improvement associated with consistently estimating $E[Y|A,X]$ suggest a potential advantage of using more flexible estimation procedures to estimate $E[Y|A=l,X]$. Replacing $h(\hat \beta_0 + \hat \beta_1 A + \hat \beta_2^T X)$ by a more data adaptive estimator for $E[Y|A,X]$ does not guarantee consistency of the model standardization estimator unless the model for $E[Y|A=l,X]$ is correctly specified. 

Equation \eqref{Var-1} in Supplementary Web Appendix \ref{sec:consistency} shows that the asymptotic variance of $\hat \mu_{(MS,l)}(w)$ depends on $Var(\hat \beta)$. This suggests that efficiency could potentially be improved by using the whole dataset to estimate the regression coefficient instead of just observations falling in node $w$. Furthermore, the CAITs that use $\hat \mu_{(MS)}(w)$ require the GLM to be re-calculated for all possible splits into child nodes. As the child node sample sizes can be small this limits the number of terms that can be included in the model. So the ability to include more terms in the model is another advantage of using an estimator for $\hat E[Y|A,X]$ implemented using the whole data set. Finally, only estimating $\hat E[Y|A,X]$ once prior to the tree building process rater than multiple times at each node as is needed for implementation of $\hat \mu_{(MS)}(w)$ substantially reduces the computational complexity of the CAIT algorithm.


A logical approach to utilize the whole dataset to estimate $E[Y|A,X]$ would be to use a GLM estimator $\hat \beta^*$ calculated using \eqref{Est-Eq-1} with $w = \mathcal{X}$. Supplementary Web Appendix \ref{sec:consistency} shows that if the GLM is correctly specified, then the model standardization estimator implemented by replacing $\hat \beta$ by $\hat \beta^*$ in equation \eqref{EstMS} consistently estimates $\mu_l(w)$. However, under mis-specification of \eqref{GLM-Mod} the model standardization estimator implemented using $\hat \beta^*$ is not guaranteed to be consistent.

Now we describe an estimator for  $\mu_l(w)$ that overcomes the aforementioned two disadvantages of the model standardization estimator. This estimator i) allows for more flexible estimation procedures	 for $E[Y|A, X]$ while remaining consistent under model mis-specification and ii) is consistent even if the estimator for $E[Y|A,X]$ is calculated using the whole dataset.

Let $\hat E[Y|A,X]$ be an estimator for $E[Y|A,X]$ that is not restricted to be from the class of GLMs or calculated only using data in group $w$. Define the covariate adjusted estimator
\begin{equation}
\label{EstDA}
\hat \mu_{(DA,l)}(w) = \hat \mu_{(Unad,l)}(w) - \frac{1}{n(w)} \sum_{i=1}^n I(X_i \in w) \frac{I(A_i = l) - \frac{n_l(w)}{n(w)}}{\frac{n_l(w)}{n(w)}} \hat E[Y|A=l,X_i].
\end{equation}
When $w = \mathcal{X}$ this is a special case of the class of estimators developed in \citet{zhang2008improving} and is further discussed in \citet{bartlett2017covariate}. By the independence of $A$ and $X$, $E[I(X \in w) (I(A = l) - n_l(w)/n(w)) \hat E[Y|A=l,X]] =0$. Hence, $\hat \mu_{(DA,l)}(w)$ is a consistent estimator for $\mu_l(w)$ even if $\hat E[Y|A,X]$ is mis-specified. When the model for the conditional expectation is correctly specified and calculated using the dataset $\mathcal{F}(w) = \{(Y_i, A_i, X_i), i \in w\}$, $\hat \mu_{(DA,l)}(w)$ is asymptotically efficient within a class of estimators satisfying the semi-parametric framework defined by equations $(8)$ and $(9)$ in \citet{zhang2008improving} calculated using $\mathcal{F}(w)$. A variance estimator for $\hat \mu_{(DA,l)}(w)$ is given by Equation \eqref{Var-2} in Supplementary Web Appendix \ref{sec:consistency}.


We refer to the CAITs with $\hat \mu_{Unad,l}(w)$ as the unadjusted CAIT algorithm, the CAIT with $\hat \mu_{(MS,l)}(w)$ as the model standardization CAIT algorithm, and the CAIT with $\hat \mu_{(DA,l)}(w)$ as the data adaptive CAIT algorithm.

\section{Simulations}
\label{sec:Simulations}

\subsection{Simulation Setup and Evaluation Measures}
\label{sec:Setup}
We will use the following two simulation settings to evaluate the performance of the CAIT algorithms.
\begin{itemize}
\item The covariate vector is simulated from a five dimensional mean zero normal distribution with $Cov(X^{(j)}, X^{(k)}) = 0.3$ for $j \neq k$ and $Var(X^{(j)}) = 1$. The treatment indicator is simulated from a $Bernoulli(0.5)$ distribution. The outcome is simulated from $Y = 2 + 2 * X^{(1)} + 2 * A * I(X^{(1)} < 0) + e^{X^{(2)}} + \varepsilon$, with $\varepsilon \sim \mathcal{N}(0,1)$. For this simulation setting the treatment effect differs depending on whether $X_1 < 0$ or not and the correct tree structure therefore splits on $X_1$ at 0. The training set consists of $500$ independent observations simulated from the joint distribution of $(Y,A,X)$ and the test set is of size $1000$. We refer to this simulation setting as the setting with heterogeneous treatment effect. 
\item The covariate vector is simulated from a five dimensional mean zero normal distribution with $Cov(X^{(j)}, X^{(k)}) = 0.3$ for $j \neq k$ and $Var(X^{(j)}) = 1$. The treatment indicator is simulated from a $Bernoulli(0.5)$ distribution. The outcome is simulated using the formula $Y = 2 + 2 * A + 2 X^{(1)} + e^{X^{(2)}} + \varepsilon$, with $\varepsilon \sim \mathcal{N}(0,1)$. For this simulation setting, the treatment effect is the same for all covariate values and the correct tree consists only of the root node. The training set consists of $500$ independent observations simulated from the joint distribution of $(Y,A,X)$ and the test set is of size $1000$. We refer to this simulation setting as the setting with homogeneous treatment effect.
\end{itemize}
To evaluate the performance of the tree building algorithms we use the following evaluation measures. 
\begin{itemize}
\item Mean Squared Error. Let $\hat \alpha(X_i)$ be the model prediction for $\alpha(X_i) = E[Y|A=1,X_i] - E[Y|A=0,X_i]$. The mean squared error is defined as $1000^{-1} \sum_{i=1}^{1000} (\hat \alpha(X_i) - \alpha(X_i))^2$, where $X_i, i = 1, \ldots, 1000$ are the covariates from the test set. 
\item Proportion of correct trees: For a continuous covariate the probability of getting exactly the correct split point is zero. Following \citet{steingrimsson2016doubly}, we define a tree to be correct if it splits on all variables the correct number of times independently of the ordering or the selection of splitting point.
\item Number of noise variables: The average number of times the tree splits on the noise variables ($X^{(1)},X^{(2)}, X^{(3)}, X^{(4)}, X^{(5)}$ for the homogeneous treatment effect setting and $X^{(2)}, X^{(3)}, X^{(4)}, X^{(5)}$ for the heterogeneous treatment effect setting).
\end{itemize}

\subsection{Implementation of Algorithms}
\label{sec:Implementation}

We implemented the large tree $\psi_{Max}$ for the CAIT algorithms using the ability of {\tt rpart} to accommodate user written splitting and evaluation functions. This allows the use of {\tt rpart's} framework and plotting options. To allow enough observations to fit the GLM, we set the {\tt minbucket} parameter to $30$ for all three CAIT algorithms and to not further restrict the size of $\psi_{Max}$ we set the {\tt cp} parameter to 0. All other tuning parameters are set as the default for the {\tt rpart} function. For the CAITs implemented using $\hat \mu_{(MS, l)}(w)$, we replace $\hat \mu_{(MS, l)}(w)$ by $\hat \mu_{(Unad, l)}(w)$ if the minimum number of observations assigned to each treatment arm in the node is less than $10$. 

The GLM used to calculate $\hat \mu_{(MS,l)}(w)$ consist of main effects of treatment and all five covariates. For implementation of $\hat \mu_{(DA,l)}(w)$, a generalized additive model is used to estimate $E[Y|A, X]$. The model includes the main effects of treatment and covariates for which a smoothing spline with three degrees of freedom is used. 

Neither the GLM used to implement $\hat \mu_{(MS,l)}(w)$ nor the generalized additive model used for $\hat \mu_{(DA,l)}(w)$ are correctly specified. For both settings, the GLM uses the wrong functional form for the covariate $X^{(2)}$ and for the heterogeneous setting it fails to include an interaction between $A$ and $I(X^{(1)} < 0)$. For the heterogeneous setting, the generalized additive model does also not include an interaction between $A$ and $I(X^{(1)} < 0)$. In addition, both models include the noise covariates $X^{(3)}, X^{(4)},$ and $X^{(5)}$.

To evaluate the impact of mis-specifying the GLM and GAMs we also implement model standardization and data adaptive CAITs with correct model specification. For the setting with heterogeneous treatment effect the linear model used for the model standardization CAIT is $\hat \beta_0 + \hat \beta_1 A + \hat \beta_2 X^{(1)} + \hat \beta_3 e^{X^{(2)}} + \hat \beta_4 A * X^{(1)}$. Hence, the linear model does not use the completely correct specification which would replace $A * X^{(1)}$ by $A * I(X^{(1)} < 0)$. But, it is closer to the correct model as it uses the correct functional form for the main effect of $X^{(2)}$, includes an interaction between $A$ and $X^{(1)}$, and does not include the noise covariates $X^{(3)},X^{(4)}$, and $X^{(5)}$. The correct GAM model allows the effect of $X^{(1)}$ to differ depending on the level of $A$ for the setting with heterogeneous treatment effect and the noise variables $X^{(3)},X^{(4)}$, and $X^{(5)}$ are not included for both simulation settings.

The random forest algorithm used for final tree selection method 2 is fit using the {\tt rfsrc} function from the {\tt randomForestSRC} package using all default tuning parameters. For all CAIT algorithms, final tree selection method 1 uses $\lambda = 4$. Final tree selection method 1 also requires splitting the training set into an initial tree building and a validation set. In the simulations the validation set is a random sample of the training set of size 100.

We compare the performance of the CAIT algorithms to both the model based recursive partitioning method (MOB) of \citet{seibold2016model} and the virtual twins (VT) method of \citet{foster2011subgroup}. The MOB algorithm was implemented with the {\tt partykit} package using the {\tt glmtree} function \citep{hothorn2015partykit}. We implemented two versions of MOB, one which all covariates are included as both main effects and in the ``tree part'' and one which the covariates are only included in the ``tree part''. MOB trees implemented by including covariates only in the ``tree part'' tended to split frequently on the variable $X^{(2)}$ which is prognostic but not predictive and the results were worse compared to also including the covariates as main effects. Hence, we omit the results for the MOB algorithm which does not include main effects of covariates. For the VT algorithm the random forest estimator included the terms $AX^{(j)}, (1-A)X^{(j)}$ for $j =1, \ldots, p$. The final tree for the VT method was fit using {\tt rpart} with the same tuning parameters as in \citet[Section 2.3.2]{foster2011subgroup}. This includes setting the minimal terminal node size to $20$ and the complexity parameter to $0.02$. Code implementing the simulations presented in this section is available from github.com/jas757/CAIT.

\subsection{Simulation Results}
\label{sec:Results}
\begin{figure}
\begin{center}
\includegraphics[scale = 0.6]{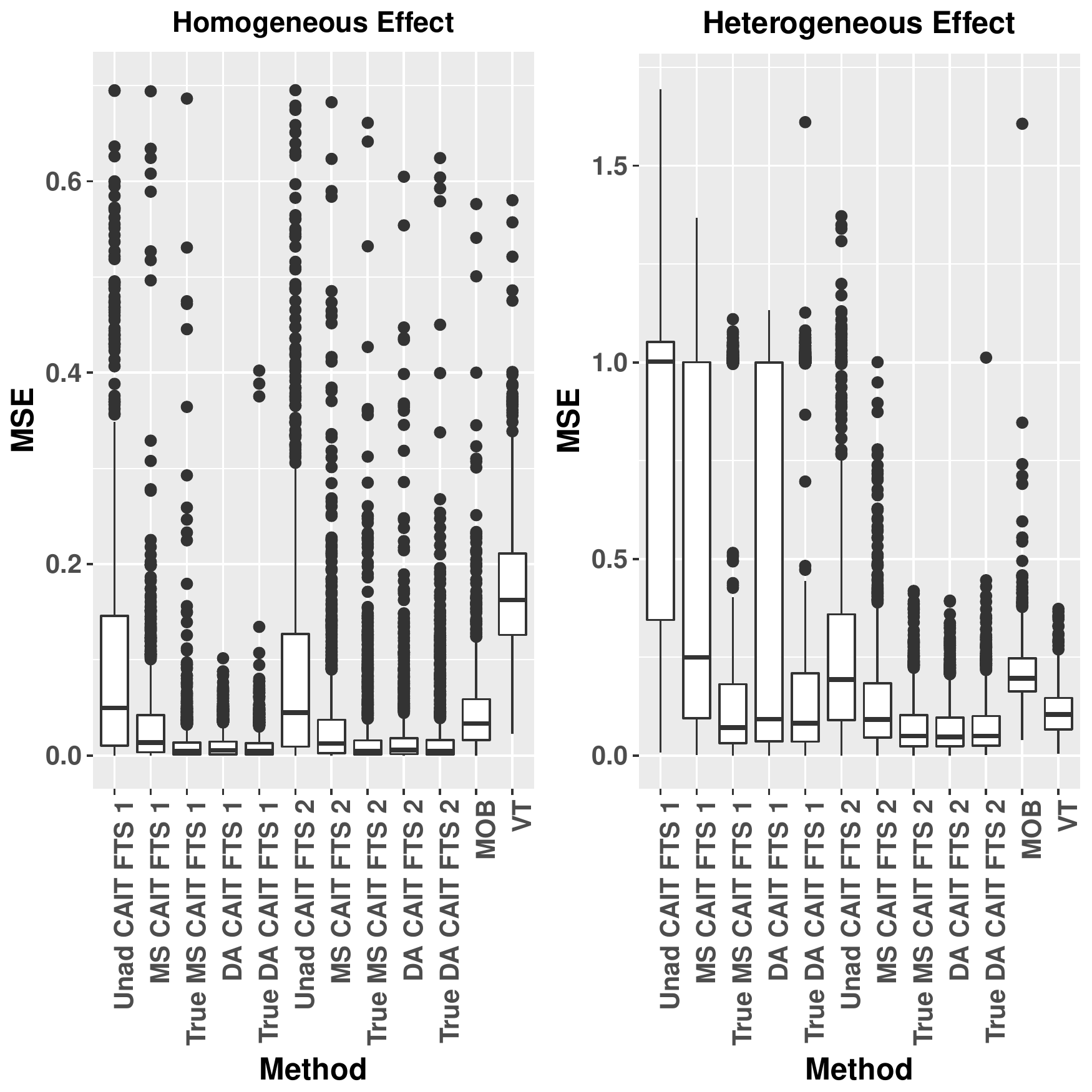} 
\caption{Mean squared error for the eight different algorithms for both simulation settings described in Section \ref{sec:Setup} with lower values indicating better performance. The left (right) plot shows simulation results when the treatment effect is homogeneous (heterogeneous). Unad CAIT refers to the unadjusted CAIT algorithm. MS CAIT and DA CAIT are the model standardization and data adaptive CAITs. True refers to that the correct GLM and GAM is used for the MS CAIT and DA CAIT. FTS 1 and FTS 2 denote if method 1 or method 2 was used for final tree selection for the CAIT algorithms. MOB is the model based recursive partitioning of \citet{seibold2016model} and VT is the virtual twins algorithm of \citet{foster2011subgroup}.}
\label{Main-MSE}
\end{center}
\end{figure}

To evaluate the performance of the different algorithms we used $1000$ simulations for both settings described in Section \ref{sec:Setup}. Figure \ref{Main-MSE} shows boxplots of MSE and Table \ref{Table-Main} shows proportion of correct trees and average number of noise variables for the CAIT algorithms, the MOB and VT methods.

\begin{table}[]
\centering
\begin{tabular}{|l|ll|ll|}
\hline
           & \multicolumn{2}{|l|}{Homogeneous Effect} & \multicolumn{2}{l|}{Heterogeneous Effect} \\
           \hline
           & Correct Trees         & Numb Noise         & Correct Trees          & Numb Noise          \\
Unad CAIT FTS 1 & 0.90              & 0.31               & 0.28               & 0.65                \\
MS CAIT FTS 1  & 0.97              & 0.046              & 0.50               & 0.33                \\
True MS CAIT FTS 1  &      0.97         &   0.042            &         0.68       &  0.21               \\
DA CAIT FTS 1  & 1.00              & 0.001              & 0.61               & 0.074               \\
True DA CAIT FTS 1 &        0.99       &    0.007          &          0.65      &   0.23             \\
Unad CAIT FTS 2 & 0.98              & 0.029               & 0.86               & 0.070               \\
MS CAIT FTS 2  & 0.92              & 0.17               & 0.82               & 0.19                \\
True MS CAIT FTS 2 &        0.88       &    0.19            &         0.80       &  0.28               \\
DA CAIT FTS 2  & 0.90              & 0.14               & 0.81               & 0.26                \\
True DA CAIT FTS 2 &        0.89       &    0.17            &         0.79       &  0.26               \\
MOB        & 0.00              & 1.9                & 0.00               & 1.13                \\
VT         & 0.00              & 5.7                & 0.16               & 1.30 \\
\hline              
\end{tabular}
\caption{Proportion of correct trees (higher is better) and average number of noise variables used for splitting (lower is better). Columns 2 and 3 show simulation results when the treatment effect is homogeneous and columns 4 and 5 show simulation results when the treatment effect is heterogeneous. Unad CAIT refers to the unadjusted CAIT algorithm. MS CAIT and DA CAIT are the model standardization and data adaptive CAITs. True refers to that the correct GLM and GAM is used for the MS CAIT and DA CAIT. FTS 1 and FTS 2 denote if method 1 or method 2 was used for final tree selection for the CAIT algorithms. MOB is the model based recursive partitioning of \citet{seibold2016model} and VT is the virtual twins algorithm of \citet{foster2011subgroup}.}
\label{Table-Main}
\end{table}

Figure \ref{Main-MSE} shows that for both simulation settings and both final tree selection methods the model standardization and data adaptive CAITs have smaller MSE than the unadjusted CAIT. This is true both when the models needed to implement the model standardization CAIT and the data adaptive CAIT are mis-specified and when they are correctly specified. This is consistent with the efficiency gains expected when using covariate adjusted estimators. In agreement with the asymptotic results, both the model standardization and data adaptive CAITs have smaller MSE when the corresponding model is correctly specified. The improvement is larger for model standardization CAIT, which is expected as the GAM model used to implement the data adaptive CAIT is more flexible and therefore closer to being correctly specified.

In the homogeneous treatment effect setting, both final tree selection methods show similar performance. In the heterogeneous treatment effect setting, final tree selection method 2 performs better than final tree selection method 1.


Both the MOB and VT methods build on average larger trees than the true tree for both settings and fit substantially more often one noise variables compared to the CAIT methods. The CAIT algorithms with $\hat \mu_{(MS)}(w)$ and $\hat \mu_{(DA)}(w)$ perform better than both MOB and VT on all evaluation measures for both settings with the exception that the CAIT algorithms combined with final tree selection method 1 and implemented using mis-specified models have higher MSE when the treatment effect is heterogeneous. This is due to the CAIT algorithms combined with final tree selection method 1 and mis-specified models sometimes under-fit, with $27\%$ and $29\%$ of the final trees consisting only of the root node for the model standardization and data adaptive CAITs, respectively. On the other hand, the MOB and VT fit too large trees with $86\%$ and $46\%$ of the final trees being of size $3$ for the MOB and VT methods, respectively. For the heterogeneous simulation setting, MSE more severely penalizes under-fitting than over-fitting. When the model standardization and data adaptive CAITs use the correct model specification, the MSE is either better or comparable to both the MOB and VT methods.

In Supplementary Web Appendix \ref{swa:DA} we present additional simulation results when the sample size is increased and for simulations where the outcome is binary. When the sample size is increased to $1000$ all methods show improved performance and the relative performance of the methods is similar to what is seen in Figure \ref{Main-MSE} and Table \ref{Table-Main}. The simulations with binary outcomes also show similar trends to what is seen in Figure \ref{Main-MSE} and Table \ref{Table-Main}.

\section{Analysis of Substance Abuse Treatment Engagement Trial}
\label{sec:Analysis}

We use the eight algorithms compared in Section \ref{sec:Results} to analyze data from a clinical trial comparing treatment engagement options for substance abusers \citep{carroll2006motivational}. At the time of submission, the dataset is publicly available at datashare.nida.nih.gov. The participants were randomized to either motivational interviews plus standard of care or to standard of care only. One of the aims of the trial was to compare the groups in terms of treatment engagement. The outcome we focus on is the number of sessions completed in the 28 days after treatment assignment. The data was prepared using the code provided in the Supplementary Material accompanying \citet{doove2014comparison}, with the exception that we combine the levels of primary drug used into alcohol and not alcohol. This is done due to few participants indicating opioid, methamphetamine, marijuana, or cocaine being their primary drug. We furthermore combine the levels of ethnicity into white and not white, and the levels of marital status into never married and is or has been married.

The dataset analyzed consists of $352$ participants and 18 covariates which are: gender, ethnicity, employment status, marital status, if admission was prompted by legal system, if the participant is on probation or parole, any previous alcohol treatment, the principal drug used, age, years of education, number of days of substance abuse in the last 30 days, and composite addiction severity index (ASI) for medical, employment, alcohol, legal, psychological, drug, and family. The standard of care arm had $178$ participants and the motivational interview plus standard of care arm had 174 participants. For further details on the study we refer to \citet{carroll2006motivational}. 

All eight algorithms were implemented as described in Section \ref{sec:Implementation}. The GLM used to implement the model standardization CAIT algorithm includes the main effects of treatment, psychological and education composite ASI. The reason for including the two latter variables is that a univariate linear model analysis shows that they are strongly prognostic. The GAM used to implement the data adaptive CAIT includes the main effects of treatment, gender, psychological and education composite ASI, number of days of substance abuse in the last 30 days, and an indicator if admission was prompted by legal system. 

The final tree for all CAIT algorithms and the MOB algorithm with the covariates also included as main effects consists only of a root node. That is, none of these algorithms make any splits. The large tree $\psi_{Max}$ built by all the CAITs first splits on if the composite ASI drug score \citep{mcgahan1986composite} measured prior to randomization is greater than $0.26$ or not. The group with drug ASI larger than $0.26$ is small consisting only of $42$ participants. The other splits for $\psi_{Max}$ differed between the different CAITs. That the final CAIT trees consist only of the root node suggest that there is no concrete evidence for any subgroup having enhanced treatment effect. The consistent first split on drug ASI of $0.26$ suggest that the treatment effect might differ depending on the drug ASI scale but the small sample size in the large drug ASI group makes that hard to infer with any certainty.

Figures \ref{VT-Tree} and \ref{Mob-Tree} in Supplementary Web Appendix \ref{swa:DA} show the final tree structures for the VT method (final tree with six terminal nodes) and the MOB method when the covariates are only included in the ``tree part'' (final tree with three terminal nodes). The larger final trees built by these methods is consistent with the trends seen in the simulations. 


\section{Discussion}
\label{sec:discussion}

This manuscript develops a new recursive partitioning method for subgroup identification which replaces unadjusted treatment effect estimators by more efficient covariate adjusted estimators. Several potentially interesting future research directions include: extensions to time to event outcomes, extensions to more complex sampling designs such as cluster randomized trials or longitudinal data structures, and extensions to different recursive partitioning methods such as partDSA \citep{molinaro2010partdsa} or GUIDE \citep{loh2015regression}.



Data adaptive methods, such as regression tree based methods, can overestimate the treatment effect in the subgroups they identify. \citet{foster2011subgroup} discussed several methods for bias correction based on either resubstitution, cross-validation, or the bootstrap. All of these methods can be used on connection with the CAIT algorithms.


\noindent

\setcounter{section}{19}
\renewcommand{\thesection}{\Alph{section}}
\newpage 
\setcounter{page}{1}

\section*{Supplementary Web Appendix}

\section{Additional Simulation Results}
\label{swa:Simulations}

\subsection{Simulations with Increased Sample Size}
\label{swa:1000-Samp}

\begin{figure}
\begin{center}
\includegraphics[scale = 0.4]{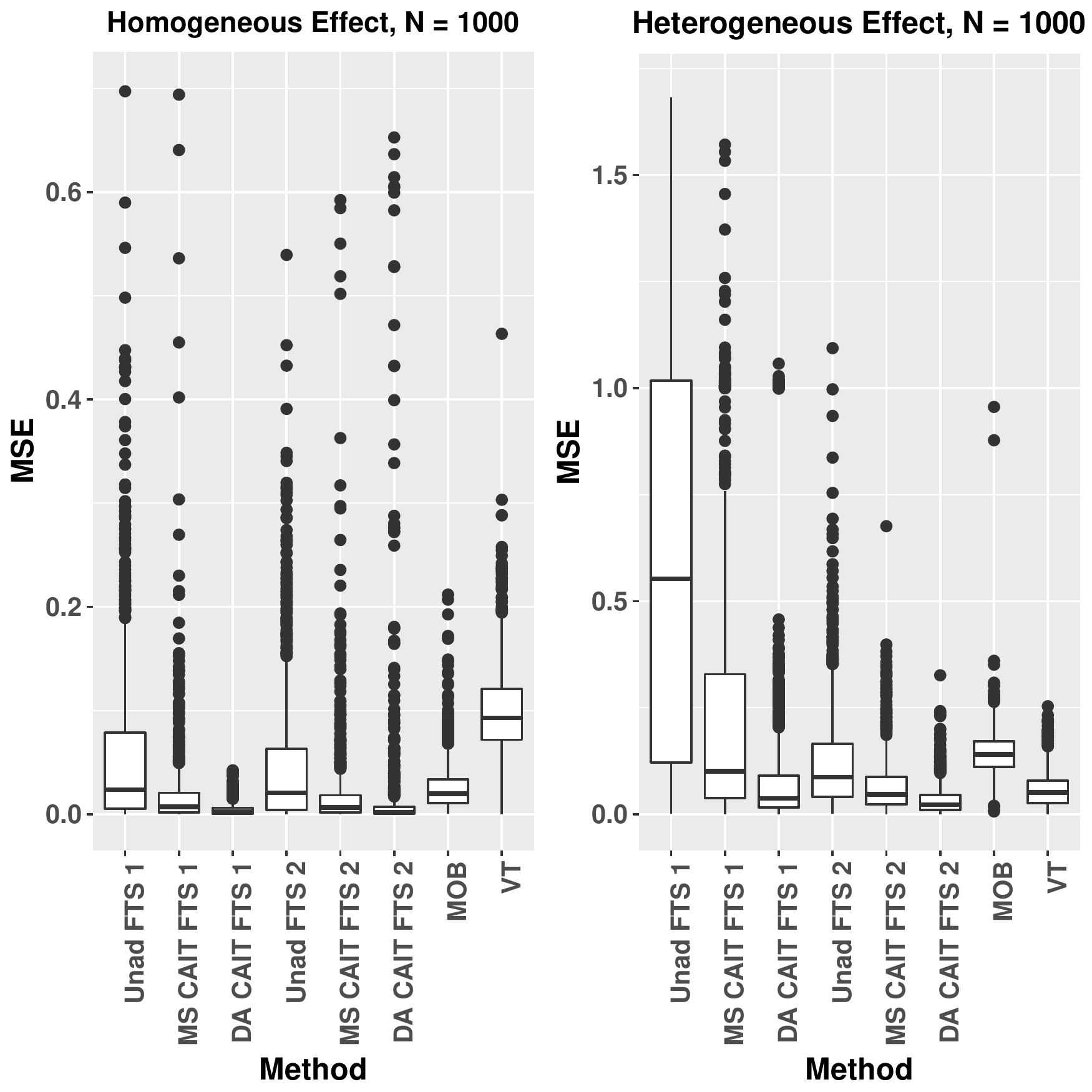} 
\caption{Mean squared error for the eight different algorithms for both simulation settings described in Section \ref{sec:Setup} when the sample size is 1000. Lower values indicate better performance. The left (right) plot shows simulation results when the treatment effect is homogeneous (heterogeneous). Unad CAIT refers to the unadjusted CAIT algorithm. MS CAIT and DA CAIT are the model standardization and data adaptive CAITs. FTS 1 and FTS 2 denote if method 1 or method 2 was used for final tree selection for the CAIT algorithms. MOB is the model based recursive partitioning of \citet{seibold2016model} and VT is the virtual twins algorithm of \citet{foster2011subgroup}.}
\label{1000Samp-MSE}
\end{center}
\end{figure}

Figure \ref{1000Samp-MSE} shows the results when the sample size for both settings described in Section \ref{sec:Setup} is increased to $1000$. For final tree selection method 1 the size of the sample used to build $\psi_{Max}$ is $800$ and the remaining $200$ observations are used as the validation sample for final tree selection. All other parameters are as described in Section \ref{sec:Implementation}. The results show that the CAIT algorithms perform better as the sample size is increased and the relative performance of the methods is similar to what is seen in Figure \ref{Main-MSE} and Table \ref{Table-Main}.

\FloatBarrier

\begin{table}[]
\centering
\begin{tabular}{|l|ll|ll|}
\hline
           & \multicolumn{2}{|l|}{Homogeneous Effect} & \multicolumn{2}{l|}{Heterogeneous Effect} \\
           \hline
           & Correct Trees         & Numb Noise         & Correct Trees          & Numb Noise          \\
Unad CAIT FTS 1 &      0.90         &   0.63             &     0.38           &    2.2             \\
MS CAIT FTS 1  &      0.97         &   0.055            &      0.58          &     1.4            \\
DA CAIT FTS 1  &      1.00        &   0.00           &        0.73        &       0.74         \\
Unad CAIT FTS 2 &      0.99         &  0.016              &     0.91           &    0.072            \\
MS CAIT FTS 2  &      0.93        &   0.14             &     0.87           &    0.13             \\
DA CAIT FTS 2  &      0.91        &   0.15             &     0.85           &    0.19             \\
MOB        &    0.00          &   2.5              &     0.00           &    2.9             \\
VT         &    0.004           &   5.4             &    0.36            &  0.77 \\
\hline              
\end{tabular}
\caption{Proportion of correct trees (higher is better) and average number of noise variables used for splitting (lower is better). The sample size is 1000. Columns 2 and 3 show simulation results when the treatment effect is homogeneous and columns 4 and 5 show simulation results when the treatment effect is heterogeneous. Unad CAIT refers to the unadjusted CAIT algorithm. MS CAIT and DA CAIT are the model standardization and data adaptive CAITs. FTS 1 and FTS 2 denote if method 1 or method 2 was used for final tree selection for the CAIT algorithms. MOB is the model based recursive partitioning of \citet{seibold2016model} and VT is the virtual twins algorithm of \citet{foster2011subgroup}.}
\label{Table-1000Samp}
\end{table}

\subsection{Simulations for a Binary Outcome}

This section presents simulation results for a binary outcome. As in the main simulations we use two simulation settings, one with a heterogeneous treatment effect and one with a homogeneous treatment effect. The settings are:
\begin{itemize}
\item The covariate vector is simulated from a five dimensional mean zero normal distribution with $Cov(X^{(j)}, X^{(k)}) = 0.3$ for $j \neq k$ and $Var(X^{(j)}) = 1$. The treatment indicator is simulated from a $Bernoulli(0.5)$ distribution. The outcome is simulated from a Bernoulli distribution with $P(Y = 1|A, X) = 0.1 + 0.3 A I(X^{(1)} < 0) + 0.3 * e^{X^{(2)}}/(1+e^{X^{(2)}})$. Here the treatment effect differs depending on if $X^{(1)} < 0$, and the correct tree therefore makes a single split at $X^{(1)} = 0$. The training set consists of $1000$ independent observations simulated from the joint distribution of $(Y,A,X)$ and the test set is of size $1000$. We refer to this simulation setting as the setting with heterogeneous treatment effect. 
\item The covariate vector is simulated from a five dimensional mean zero normal distribution with $Cov(X^{(j)}, X^{(k)}) = 0.3$ for $j \neq k$ and $Var(X^{(j)}) = 1$. The treatment indicator is simulated from a $Bernoulli(0.5)$ distribution. The outcome is simulated from a Bernoulli distribution with $P(Y = 1|A, X) = 0.1 + 0.3 * e^{X^{(2)}}/(1+e^{X^{(2)}})$. For this setting the treatment effect is the same for all values of the covariate vector and the correct tree consists only of the root node. The training set consists of $1000$ independent observations simulated from the joint distribution of $(Y,A,X)$ and the test set is of size $1000$. We refer to this simulation setting as the setting with homogeneous treatment effect. 
\end{itemize}
For both simulation settings the GLM needed to implement the model standardization CAIT with is a logistic regression model with main effect of $A,X^{(1)},X^{(2)},X^{(3)},X^{(4)}$, and $X^{(5)}$. For both simulation settings the GAM needed to implement the data adaptive CAIT consists of main effects of treatment and all covariates where the covariates are modeled using smoothing splines with three degrees of freedom. For both settings both models are incorrectly specified.

The validation sample needed to implement final tree selection method 1 is of size 200 and the penalization parameter $\lambda$ is set to $4$. All other parameters are set as in Section \ref{sec:Implementation}.

\begin{figure}
\begin{center}
\includegraphics[scale = 0.4]{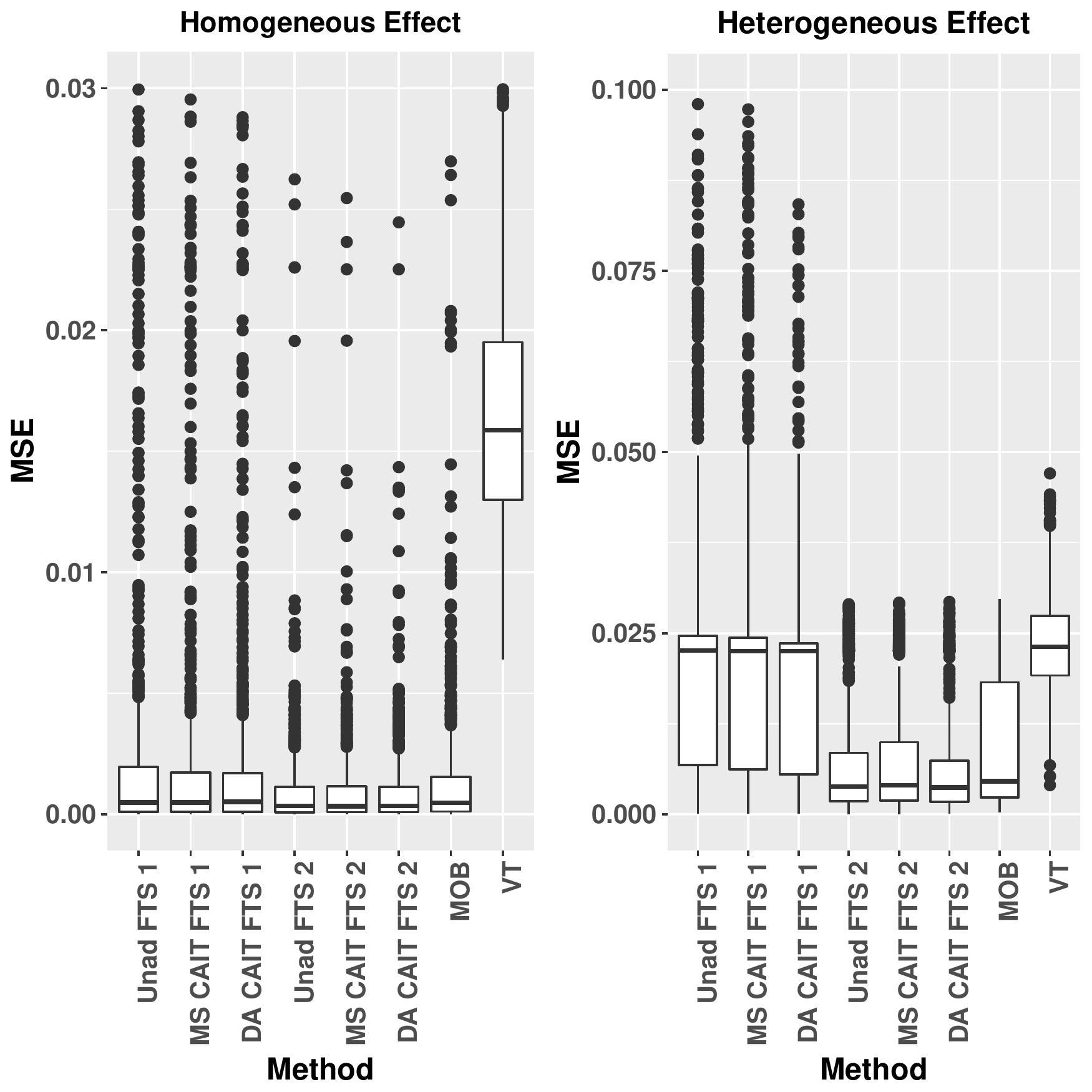} 
\caption{Mean squared error for the eight different algorithms when the outcome is binary. Lower values indicate better performance. The left (right) plot shows simulation results when the treatment effect is homogeneous (heterogeneous). Unad CAIT refers to the unadjusted CAIT algorithm. MS CAIT and DA CAIT are the model standardization and data adaptive CAITs. FTS 1 and FTS 2 denote if method 1 or method 2 was used for final tree selection for the CAIT algorithms. MOB is the model based recursive partitioning of \citet{seibold2016model} and VT is the virtual twins algorithm of \citet{foster2011subgroup}.}
\label{Bin-MSE}
\end{center}
\end{figure}

Figure \ref{Bin-MSE} shows boxplots of MSE for the CAITs and the MOB and VT methods. Table \ref{Table-Bin} shows number of correct trees and average number of splits on noise variables. The results show similar trends to the main simulations presented in Section \ref{sec:Results}. The difference in performance between the CAIT algorithms is smaller than for a continuous outcome. 

\begin{table}[]
\centering
\begin{tabular}{|l|ll|ll|}
\hline
           & \multicolumn{2}{|l|}{Homogeneous Effect} & \multicolumn{2}{l|}{Heterogeneous Effect} \\
           \hline
           & Correct Trees         & Numb Noise         & Correct Trees          & Numb Noise          \\
Unad CAIT FTS 1 &     0.85      &    0.59       &      0.33          &    0.88             \\
MS CAIT FTS 1  &     0.89      &    0.34           &       0.36        &      0.92           \\
DA CAIT FTS 1  &     0.91      &    0.15          &       0.38        &      0.35          \\
Unad CAIT FTS 2 &     1.00      &    0.001            &      0.85       &       0.032         \\
MS CAIT FTS 2  &     0.99      &    0.008            &       0.81         &     0.030            \\
DA CAIT FTS 2  &     1.00      &    0.001               &      0.88          &    0.035             \\
MOB        &     0.84      &    0.17             &     0.67          &    0.12             \\
VT         &     0.71      &    0.81             &     0.01           &  3.2 \\
\hline              
\end{tabular}
\caption{Proportion of correct trees (higher is better) and average number of noise variables used for splitting (lower is better) when the outcome is binary. Columns 2 and 3 show simulation results when the treatment effect is homogeneous and columns 4 and 5 show simulation results when the treatment effect is heterogeneous. Unad CAIT refers to the unadjusted CAIT algorithm. MS CAIT and DA CAIT are the model standardization and data adaptive CAITs. FTS 1 and FTS 2 denote if method 1 or method 2 was used for final tree selection for the CAIT algorithms. MOB is the model based recursive partitioning method of \citet{seibold2016model} and VT is the virtual twins algorithm of \citet{foster2011subgroup}.}
\label{Table-Bin}
\end{table}

\section{Supporting Material for Data Analysis}
\label{swa:DA}

\begin{figure}
\begin{center}
\includegraphics[scale = 0.4]{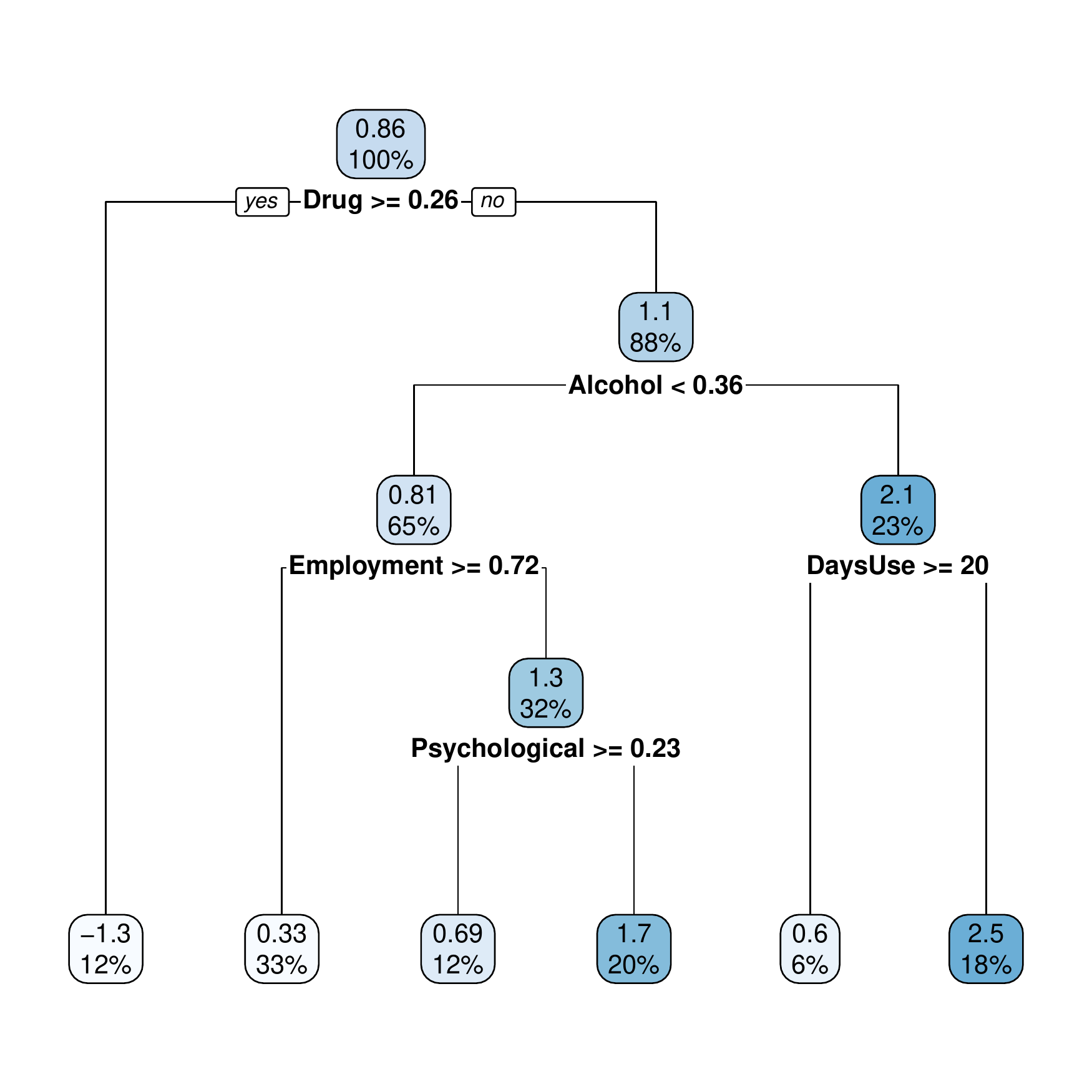} 
\caption{Final tree structure when the Virtual Twins method is applied to the substance abuse treatment engagement data from Section \ref{sec:Analysis}. Drug, Alcohol, Employment, and Psychological are the composite addiction severity indexes for drug, alcohol, employment, and psychological, respectively. DaysUse is the number of days of substance use in the past 30 days.}
\label{VT-Tree}
\end{center}
\end{figure}

\begin{figure}
\begin{center}
\includegraphics[scale = 0.4]{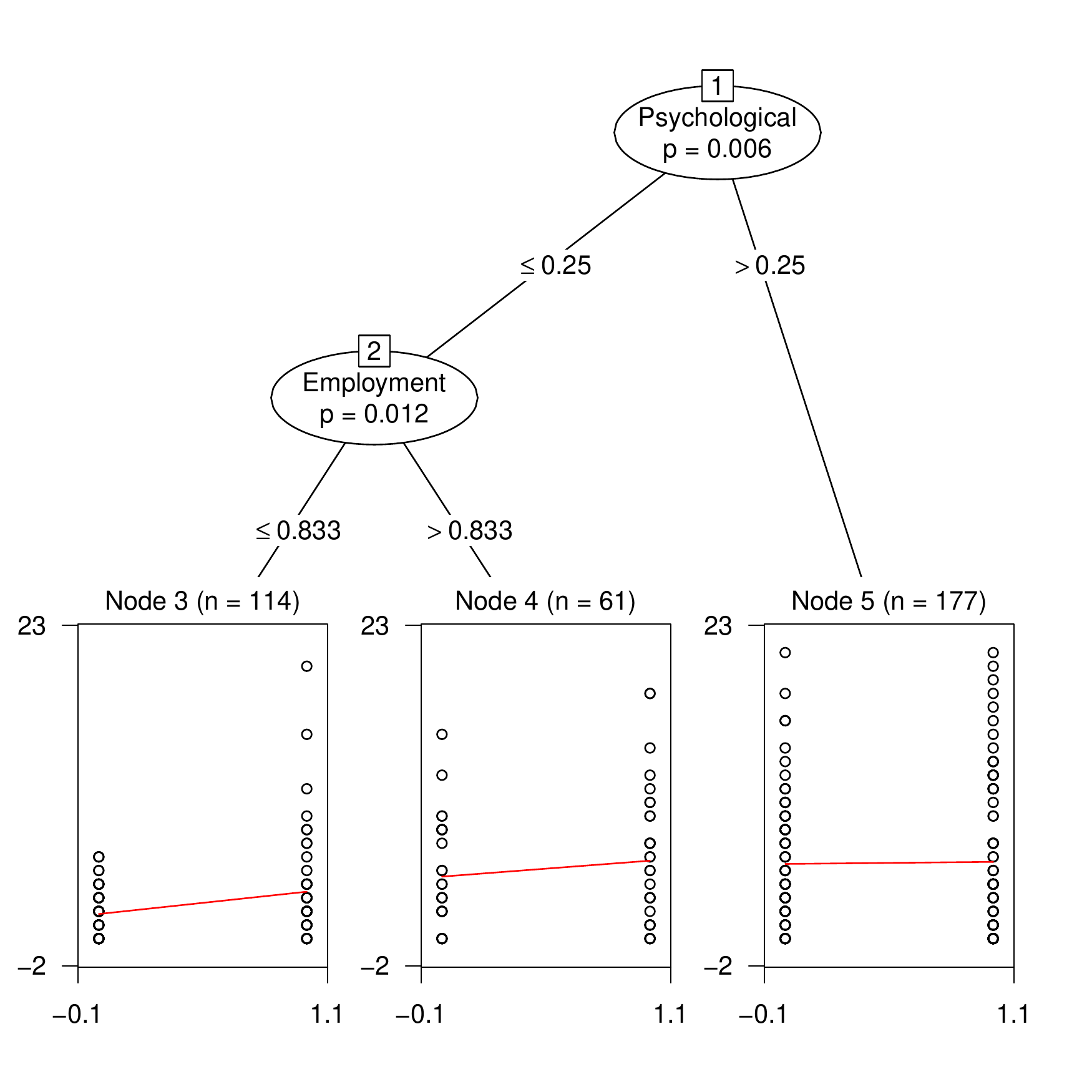} 
\caption{Final tree structure when the MOB method which does not include main effects for covariates is applied to the substance abuse treatment engagement data from Section \ref{sec:Analysis}. Employment, and Psychological are the composite addiction severity indexes for employment, and psychological, respectively.}
\label{Mob-Tree}
\end{center}
\end{figure}

Figures \ref{VT-Tree} and \ref{Mob-Tree} show the final tree structures for the VT method and the MOB method when main effects are not included. The simulations showed that the MOB tree without main effects tended to split on covariates that were prognostic but not necessarily predictive. This might also be the case here as both employment and psychological composite addiction severity indexes are found to be highly prognostic in a univariate analysis.

\FloatBarrier

\section{Derivations of Properties of the Covariate Adjusted Node Specific Estimators}
\label{sec:consistency}

\textbf{Consistency of $\hat \mu_{(MS,l)}(w)$:} Recall that the asymptotic limit of $\hat \beta$, denoted by $\tilde \beta$, satisfies the equation 
\[
E[I(X \in w) (Y - h(\tilde \beta_0 + \tilde \beta_1 A + \tilde \beta_2^T X)) (1, A, X^T)^T] = 0.
\]
The first element of the estimating equation and that $A$ is independent of $X$ gives
\begin{align}
E[I(X \in w)Y] &= E[I(X \in w) h(\tilde \beta_0 + \tilde \beta_1 A + \tilde \beta_2^T X)] \nonumber \\ 
&= 0.5 E[I(X \in w) h(\tilde \beta_0 + \tilde \beta_1 A + \tilde \beta_2^T X)|A= 1] \nonumber \\ &+ 0.5 E[I(X \in w) h(\tilde \beta_0 + \tilde \beta_1 A + \tilde \beta_2^T X)|A= 0] \nonumber \\
& = 0.5 E[I(X \in w) h(\tilde \beta_0 + \tilde \beta_1 + \tilde \beta_2^T X)] \nonumber \\ &+ 0.5 E[I(X \in w) h(\tilde \beta_0 + \tilde \beta_2^T X)]. \label{eq-1-cons}
\end{align}
And the second element of the estimating equation gives 
\begin{align}
E[I(X \in w)YA] &= E[I(X \in w) A h(\tilde \beta_0 + \tilde \beta_1 A + \tilde \beta_2^T X)] \nonumber \\ 
&= 0.5 E[I(X \in w) h(\tilde \beta_0 + \tilde \beta_1 A + \tilde \beta_2^T X)|A= 1] \nonumber \\ 
& = 0.5 E[I(X \in w) h(\tilde \beta_0 + \tilde \beta_1 + \tilde \beta_2^T X)]. \label{eq-2-cons}
\end{align}
Using that $E[I(X \in w)Y] = E[I(X \in w)YA] + E[I(X \in w)Y(1 - A)]$, we have $E[I(X \in w)Y(1 - A)] = 0.5 E[I(X \in w) h(\tilde \beta_0 + \tilde \beta_2^T X)]$. Hence, 
\begin{align*}
E[I(X \in w)Y|A=1] &= 2 * 0.5 * E[I(X \in w) h(\tilde \beta_0 + \tilde \beta_1 A + \tilde \beta_2^T X)|A= 1] \\ 
&= E[I(X \in w) h(\tilde \beta_0 + \tilde \beta_1 + \tilde \beta_2^T X)].
\end{align*}
Similarly, $E[I(X \in w)Y|A=0] = E[I(X \in w) h(\tilde \beta_0 + \tilde \beta_2^T X)]$. Completing the proof of the consistency of $\hat \mu_{(MS,l)}(w), l = 0,1$.
\\ 
\\
\noindent \textbf{Properties of Covariate Adjusted Estimator using $\hat \beta^*$:} The estimator $\hat \beta^*$ estimated using all observations, not just the observations falling in node $w$, is calculated by solving
\begin{equation}
\label{GLM-EE-1}
\sum_{i=1}^n (Y_i - h(\beta_0^{*} + \beta_1^{*} A_i + (\beta_2^{*})^T X_i)) (1, A_i, X_i^T)^T = 0.
\end{equation}
The estimator $\hat \beta^*$ estimates the population parameter $\tilde \beta^*$ satisfying
\[
E[(Y - h(\tilde \beta_0^{*} + \tilde \beta_1^{*} A + (\tilde \beta_2^{*})^T X)) (1,A, X^T)^T] = 0.
\] 
Define the covariate adjusted estimator for $\mu_l(w)$ utilizing $\hat \beta^*$ instead $\hat \beta$ as
\begin{equation}
\label{cov-ad-star}
\hat \mu_l^*(w) = \frac{1}{n(w)} \sum_{i=1}^n I(X_i \in w) h(\hat \beta_0^{*} + \hat \beta_1^{*} l  + (\hat \beta_2^{*})^T X_i)).
\end{equation}
The quantity $h(\hat \beta_0^{*} + \hat \beta_1^{*} l + (\hat \beta_2^{*})^T X_i)$ is the prediction for an participant with treatment and covariate information $(A_i = l, X_i^T)^T$. So equation \ref{cov-ad-star} sums over the predictions for all participants falling in group $w$ setting their treatment assignment to $l$. 

Now we show that $\hat \mu_{(MS, l)}(w)$ is a consistent estimator for $\mu_{l}(w)$ if the model \eqref{GLM-Mod} is correctly specified. Under that assumption, we have
\begin{equation}
\label{Eq-used-cons}
E[h(\tilde \beta_0 + \tilde \beta_1 l + \tilde \beta_2^T X)|X \in w] = E[E[Y|X, A = l]|X \in w]= E[Y|A =l|X \in w].
\end{equation}
Hence,
\[
\lim_{n \rightarrow \infty} \hat \mu_l^*(w) = E[E[Y|X, A= l]|X \in w] = E[Y|X \in w, A =l] = \mu_l(w).
\]
The first equality sign in \eqref{Eq-used-cons} relies on \eqref{GLM-Mod} being correctly specified so the derivation above does not hold if \eqref{GLM-Mod} is misspecified.
\\ 
\\ 
\textbf{Variance Estimators for $\hat \mu_{(MS,l)}(w)$ and $\hat \mu_{(DA,l)}(w)$.} 

Define
\[
G(\beta) = \frac{1}{n(w)} \sum_{i=1}^n I(X_i \in w) \frac{\partial h(\beta_0 + \beta_1 l + \beta_2^T X_i)}{\partial \beta^T}.
\]
The asymptotic variance of the covariate adjusted estimator $\hat \mu_{(MS,l)}(w)$ can be estimated using 
\begin{equation}
\label{Var-1}
G(\hat \beta) \widehat{Var}(\hat \beta) G(\hat \beta)^T + \frac{1}{n(w)^2} \sum_{i=1}^n I(X_i \in w) \left(h(\beta_0 + \beta_1 l + \beta_2^T X_i) - \hat \mu_{(MS,l)}(w)\right)^2,
\end{equation}
where $\widehat{Var}(\hat \beta)$ is some estimator of the variance of $\widehat{Var}(\hat \beta)$. In order to guard against mis-specification of the model \eqref{GLM-Mod} a robust variance estimator, such as estimators based on the non-parametric bootstrap or robust sandwich variance estimators \citet{huber1967behavior}, needs to be used. In the simulations we use a robust variance estimator.

As derived in \citet{bartlett2017covariate} the asymptotic variance of the covariate adjusted estimator $\hat \mu_{(DA,l)}(w)$ can be estimated using
\begin{align}
&\frac{1}{n_l(w)^2} \sum_{i=1}^n I(X_i \in w) \Bigg(I(A_i = l)\bigg(Y_i - \hat \mu_{(DA,l)}(w)\bigg) \nonumber 
\\ &- \bigg(I(A_i = l) - \frac{n_l(w)}{n(w)} \bigg) \bigg(\hat E[Y|A=l, X] - \bar E[Y|A=l, X]\bigg) \Bigg)^2. \label{Var-2},
\end{align}
where $E_X[E[Y|A=l, X]]$ is an estimator for $E_X[E[Y|A=l, X]]$. Here, the outer expectation is taken w.r.t. the distribution of the covariate vector $X$. In the simulations and data analysis presented in Section \ref{sec:Simulations} and \ref{sec:Analysis} we use $E_X[E[Y|A=l, X]] = \frac{1}{n(w)}\sum_{i=1}^n I(X_i \in w)\hat E[Y|A=l, X]$. 

\end{document}